\documentclass[twoside]{article}
\usepackage[latin1]{inputenc}
\usepackage{t1enc}
\usepackage{a4}
\usepackage{tabularx}
\usepackage{epsf}
\usepackage{psfig}
\usepackage{rotate}

\def\ref{\vspace{4pt}\noindent\hangindent=10mm}
\def\arcsec{$^{''}$}
\def\arcmin{$^{'}$}

\begin{document}

\setcounter{figure}{0}
\setcounter{section}{0}
\setcounter{equation}{0}

\begin{center}
{\Large\bf
Extragalactic Globular Cluster Systems}\\[0.2cm]
{\large\bf A new Perspective on Galaxy Formation and Evolution}\\[0.7cm]

Markus Kissler-Patig \\[0.17cm]
European Southern Observatory \\
Karl-Schwarzschild-Str.~2, 85748 Garching, Germany \\
mkissler@eso.org, http://www.eso.org/$\sim$mkissler
\end{center}

\vspace{0.5cm}

\begin{abstract}
\noindent{\it
We present an overview of observational progress in the study of
extragalactic globular cluster systems. Globular clusters turn out to be 
excellent tracers not only for the star-formation histories in galaxies, but 
also for kinematics at large galactocentric radii. Their properties can be used
to efficiently constrain galaxy formation and evolution. After a brief 
introduction of the current methods and futures perspectives, we
summarize the knowledge gained in various areas of galaxy research through
the study of globular clusters. 
In particular, we address the star-formation histories of early-type
galaxies; globular cluster population in late-type galaxies and their
link to early-type galaxies; star and cluster formation during mergers and 
violent interactions; and the kinematics at large radii in early-type
galaxies. The different points are reviewed within the 
context of galaxy formation and evolution.

Finally, we revisit the globular cluster luminosity function
as a distance indicator. Despite its low popularity in the literature, we
demonstrate that it ranks among one of the most precise distance indicators 
to early-type galaxies, provided that it is applied properly.
}
\end{abstract}

\section{Preamble}

This brief review on extragalactic globular cluster systems is derived
from a lecture given for the award of the Ludwig-Bierman-Preis of the
Astronomische Gesellschaft in G\"ottingen during September 1999. 
The Oral version aimed at introducing, mostly from an observer's point of view, 
this field of research and at emphasizing its tight links to galaxy formation 
and evolution. 

The scope of this written follow-up is {\it not} to give a complete review on 
globular cluster systems but to present recent discoveries, including
examples, and to set them into the context of galaxy formation and
evolution. The choice of examples and the emphasis of certain ideas will
necessarily be subjective, and we apologize at this point for any missing
references.

Excellent recent reviews can be found in the form of two books: ``Globular
Cluster Systems'' by Ashman \& Zepf (1998), as well as ``Globular Cluster
Systems'' by Harris (2000). These include a full description of the
globular clusters in the Local group (not discussed here), as well as an 
extensive list of references, including to older reviews.

The plan of the article is the following. In section 2, we give an introduction
and the motivation for studying globular cluster systems with the aim of  
understanding galaxies.
Section 3 presents current and future methods of observations, and the
rational behind them. This section reviews recent progress in optical and 
near-infrared photometry and multi-object, low-resolution spectroscopy. 
It can be skipped by readers interested in results rather than methods.
In Section 4 we present in turn the status of our knowledge on
globular cluster systems in ellipticals, spirals and mergers.
What are the properties of the systems? How are the galaxy
types linked? And do mergers produce real `globular' clusters?
In section 5, we discuss sub-populations of globular clusters and their
possible origin. The most popular scenarios to explain the presence of
globular cluster sub-populations around galaxies are listed. The pros
and cons, as well as the expectations of each scenario are discussed.
We present, in section 6, some results from the study of globular cluster
system kinematics. 
Finally, in section 7, we revisit the globular cluster luminosity
function as a distance indicator. Under which conditions can it be used,
and how should it be applied to minimize any systematic errors? It is
compared to other distance indicators and shown to do very well.
Some conclusions and an outlook are given in section 8.


\section{Motivation}

As a reminder, globular clusters are tyically composed of $10^4$ to
$10^6$ stars clustered within a few parsecs. They are old, although
young globular-cluster-like objects are seen in mergers, and their
metallicity can vary between [Fe/H]$\simeq -2.5$ dex and [Fe/H]$>0.5$ dex.
We refer to a globular cluster system as the totallity of globular
clusters surrounding a galaxy.

\subsection{Why study globular cluster systems?}

\begin{figure}[t]
\psfig{figure=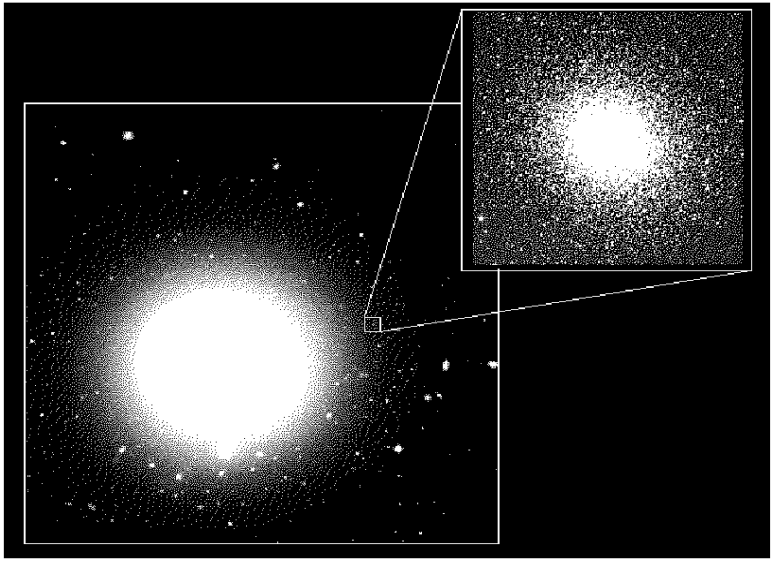,height=8cm,width=12cm
    ,bbllx=5mm,bblly=5mm,bburx=205mm,bbury=285mm,angle=-90}
\caption{The figures shows a montage of what we are observing. The
galaxy NGC 1399 of the Fornax galaxy cluster is shown on the left. It is
located at a distance of $\sim 18$ Mpc from the Milky Way. Many of the
point source surrounding it are globular clusters. If we could resolve
these clusters into stars (which we cannot), we would see clusters
similar to the Galactic clusters M~15, shown on the right.}
\end{figure}

The two fundamental questions in galaxy formation and evolution are: {\it
1)} When and how did the galaxies assemble? and {\it 2)} When and how
did the galaxies form their stars? A third question could be whether,
and to what extend, the two first points are linked.

Generally speaking, in order to answer these questions from an
observational point of view, one can follow two paths. The first would
be to observe the galaxies at high redshift, right at the epoch of
their assembly and/or star formation. We will consider this to be the
``hard way''. These observations are extremely challenging for many
reasons (shifted restframe wavelength, faint magnitudes, small angular
scales, etc...). Nevertheless, they are pursued by a number of
groups through the observations of absorption line systems along the line of 
sights of quasars, or the detection of high-redshift (e.g.~Lyman break) 
galaxies, etc... (see e.g.~Combes, Mamon, \& Charmandaris 1999, Bunker \&
van Breugel 1999, Mazure, Le Fevre, \& Le Brun 1999 for
recent proceedings on the rapidly evolving subject).

The second path is to wait until a galaxy reaches very low redshift and
try to extract information about its past. This would be the ``lazy
way''. This is partly done by the
study of the diffuse stellar populations at 0 redshift and the
comparison of its properties at low redshift. Such studies on
fundamental relations (fundamental plane, Mg-$\sigma$, Dn-$\sigma$) tend
to be consistent with the stellar populations evolving purely passively 
and having formed at high redshifts ($z>2$). 
Alternatively, one can study merging events among galaxies at
low- to intermediate-redshift (e.g.~van Dokkum et al.~1999) in order to
understand the assembly of galaxies.

How do globular clusters fit into these picture? Globular cluster
studies could be classified as the ``very lazy way'', since they reach
out to at most redshifts of $z=0.03$. However, globular clusters are
among the oldest objects in the universe, i.e.~they witnessed most,
if not all, of the history of their host galaxy. The goal of the
globular cluster system studies is therefore to extract the memory of
the system. Photometry and spectroscopy are used to derive their ages
and chemical abundances which are used to understand the epoch(s) of
star formation in the galaxy. Kinematic information obtained from the
globular clusters (especially at large galactocentric radii) can be used to 
help understanding the assembly mechanism of the galaxies. 

\subsection{Advantages of using globular clusters}

Figure 1 illustrates extragalactic globular cluster studies. We show the
galaxy NGC 1399 surrounded by a number of point sources. If these point
sources could be resolved, they would look like one of the Galactic
globular clusters (here shown as M~15). However, even with diffraction limited
imaging from space, we cannot resolve clusters at distances of 10 to 100
Mpc into single stars, and have to study their integrated properties.
The study of a globular cluster system is therefore equivalent to studying the
integrated properties of a large number of globular clusters surrounding a 
galaxy, in order to derive their individual properties and compare them,
as well as the properties of the system as a whole, with the properties of 
the host galaxy.

The purely practical advantages of observing objects at $z=0$ 
is the possibility to study the objects in great
detail. Very low $z$ observations are justified if the gain in
details outbalances the fact that at high $z$ one is seeing events closer
to the time at which they actually happened. One example that
demonstrates that the gain is significant in the case of extragalactic globular
clusters, is the discovery of several sub-populations of clusters around a 
large number of early-type galaxies. The presence of two or more distinct star-formation
epochs/mechanisms in at least a large number if not all giant galaxies
was not discovered by any other type of observations.

The old age of globular clusters is often advanced as argument for their study, 
since they witnessed the entire past of the galaxy including
the earliest epochs. If this would be the whole truth, globular clusters
would not be suited to study the recent star formation epochs. Nor would
they present a real advantage over stars, which can be old too.
What are the advantages of observing globular clusters as tracers of the
star-formation / stellar populations instead of studying directly the diffuse 
stellar population of the host galaxies?

\vskip 3mm
\noindent {\it Globular clusters trace star formation}

A number of arguments support the fact that globular clusters indeed 
trace the star formation in galaxies. However, we know that some star formation
can occur without forming globular clusters. One example is the Large
Magellanic Cloud which, at some epochs, produced stars but no clusters
(e.g.~Geha et al.~1998). On the other hand, we know that major star formation
episodes induce the formation of a large number of star clusters. For
example, the violent star formation in interacting galaxies is accompanied by
the formation of massive young star clusters (e.g.~Schweizer 1997).
Also, the final number of globular clusters in a galaxy is roughly
proportional to the galaxy luminosity, i.e.~number of stars (see Harris
1991). This hints at a close link between star and cluster
formation. Additional support for such a link comes from the close relation
between the number of young star clusters in spirals and their current star 
formation rate (Larsen \& Richtler 1999).

In summary, globular clusters are not perfect tracers for star
formation, as they will not form during every single little (i.e. low
rate) star formation event. But they will trace the major (violent)
epochs of star formation, which is our goal.

\vskip 3mm
\noindent {\it From a practical point of view}

Globular clusters exist around all luminous ($M_V>17$) 
galaxies observed to date. Their number, that scales with the mass of the
galaxy, typically lies around a few hundreds to a few thousands.

Furthermore, globular clusters can be observed out to $\sim$
100 Mpc. This is not as far as the diffuse light can be observed, but
far enough to include many thousands of galaxies of all types and in all
varieties of environments.

The study of globular clusters is therefore not restricted to a specific
type or environment of galaxies: unbiased samples can be constructed.
From this point of view, diffuse stellar light and globular clusters are 
equally appropriate.

\vskip 3mm
\noindent {\it The advantages of globular clusters over the diffuse stellar
population}

Globular clusters present a significant advantage
when trying to determine the star formation history of a galaxy: they are far
simpler structures. A globular cluster can be characterized by a single age 
and single
metallicity, while the diffuse stellar population of a galaxy needs to be
modeled by an unknown mix of ages and of metallicities. Studying a
globular cluster system returns a large number of discrete age/metallicity 
data points. These can be grouped to determine the mean ages and chemical
abundances of the main sub-populations present in the galaxy.

Along the same line, and as shown above, globular clusters form
proportionally to the number of stars. That is, the number of globular
clusters in a given population reflects the importance of the star
formation episode at its origin. Counting globular cluster in different
sub-populations indicates right away the relative importance of the
different star formation events. In contrast, the different populations
in the diffuse stellar light appear luminosity weighted: a small (in
terms of mass) but recent star formation event can outshine a much more
important but older event that has faded. 

\vskip 3mm
\noindent {\it The bonus}

As for stellar populations, kinematical informations can be derived from
the spectra originally aimed at determining ages and metallicities. Globular
clusters have the advantage that they can be traced out to
galactocentric distances unreachable with the diffuse stellar light.
The dynamical information of the clusters can be used to study the
assembly of the host galaxy. In Section 6, we will come back to this
point. 

\vskip 1cm

The bottom line is that globular clusters are good tracers for the star
formation history of their host galaxies, and eventually allow some
insight into their assembly too. They present a number of advantages
over the study of stellar populations, and complement observations at
high redshift. Their study allows new insight into galaxy formation and
evolution.


\section{Current and future observational methods}

This section is intended to give a feeling for the observational methods
used to study globular cluster systems. It addresses the problems still
encountered in imaging and spectroscopy, as well as the improvements to
be expected with future instruments. 

To set the stage: we are trying to analyze the light of objects with
typical half-light radii of 1 to 5 pc, at distances of 10 to 100 Mpc
(i.e.~half-light radii of 0.01\arcsec\ to 0.1\arcsec\ ), and absolute
magnitudes ranging from $M_V\sim -10$ to $-4$ (i.e.~$V>20$). The
galaxies in the nearest galaxy clusters (Virgo, Fornax) have globular
cluster luminosity functions that peak in magnitude around $V\sim 24$,
and globular clusters have typical half-light radii of $\sim$ 0.05\arcsec\ .

We intend to study both the properties of the individual clusters, as
well as the ones of the whole cluster system. For the individual
globular clusters, our goal is to derive their ages, chemical
abundances, sizes and eventually masses. This requires spectral
information (the crudest being just a color) and high angular resolution.
For the whole system, our goal is to determine the total numbers, the
globular cluster luminosity function, the spatial
distributions (extent or density profile, ellipticity), and any radial 
dependencies
of the cluster properties (e.g.~metallicity gradients). These properties
should also be measured for individual sub-populations, if they are present.
The requirements for the systems are therefore deep, wide-field imaging, and
the ability to distinguish potential sub-populations from each other.

\subsection{Optical photometry}

Globular clusters outside the Local Group were, for a long time, exclusively 
studied with optical photometry. Optical, ground-based photometry
(reaching $V>24$ in a field $>$ 5\arcmin $\times$ 5\arcmin\ ) turns out to be
sufficient to determine most morphological properties of the systems
(see Sect.~4).
The depth allows to reasonably sample the luminosity function, and a
field of several arcminutes a side usually covers the vast majority of the 
system.

Problems with optical photometry arise when trying to determine ages and
metallicities. It is well known that broad-band optical colors are
degenerate in age and metallicity (e.g.~Worthey 1994). A younger age is 
compensated by a higher metallicity in most broad-band, optical colors.
To some extent the problem is solved by the fact that most globular
clusters are older than several Gyr, and colors do not depend
significantly on age in that range. This, of course, means that deriving
ages from optical colors is hopeless, except for young clusters as
seen e.g.~in mergers.

For old clusters, the goal is to find a color that is as sensitive
to metallicity as possible. The widely used $V-I$ color is the least
sensitive color to metallicity. $B-V$ and $B-I$ do better in the
Johnson-Cousins system (e.g.~Couture et al.~1991 for one of the first 
comparisons), but the mini break-through came with the use of Washington 
filters (e.g.~Geisler \& Forte 1990). These allowed the discovery of the
first multi-modal globular cluster color distributions around galaxies
(Zepf \& Ashman 1993). However, the common use of the Johnson system,
combined with large errors in the photometry at faint magnitudes, do
still not allow a clean separation of individual globular clusters
into sub-populations around most galaxies.

Another problem with ground-based imaging is that its resolution is
by far insufficient to resolve globular clusters. This prohibits the
unambiguous identification of globular clusters from foreground stars
and background galaxies. In most cases the over-density of globular
clusters around the host galaxy is sufficient to derive the general
properties of the system. Control fields should be used (although often
left out because of the lack of observing time), and statistical background
subtraction performed. The individual identification of globular
clusters became possible with WFPC2 on HST. Globular clusters appear barely
extended in WFPC2 images, which allows on the one hand to reliably
separate them from foreground stars and background galaxies, and on the
other hand to systematically study for the first time globular cluster sizes
outside the Local group (Kundu \& Whitmore 1998,  Puzia et al.~1999,
Kundu et al.~1999). The disadvantage of
WFPC2 observations is that the vast majority was carried out in $V-I$,
the least performant system in terms of metallicity sensitivity.
Furthermore, the WFPC2 has a
small field of view which biases all the analysis towards the center of the
galaxies, making it very hard to derive the global properties of a
system without large extrapolations or multiple pointings.

In summary, ground-based photometry returns the general properties of
the systems, and eventually of the sub-systems when high quality
photometry is obtained. It suffers from confusion when identifying individual
clusters, and is limited in age/metallicity determinations. Space photometry 
is currently as bad in deriving ages/metallicities, but allows to
determine sizes of individual clusters. The current small fields,
however, restrict the studies of whole systems.

Future progress is expected with the many wide-field imagers coming
online, provided that deep enough photometry is obtained (errors $<0.05$
mag at $V=24$). These will provide a large number of targets for
spectroscopic follow-up. In space, the ACS to be mounted on HST will
superseed the WFPC2. The field of view remains modest, but the slightly
higher resolution will support further size determinations, and the
higher throughput will allow a more clever choice of filters, including
$U$ and $B$.

\begin{figure}[t]
\psfig{figure=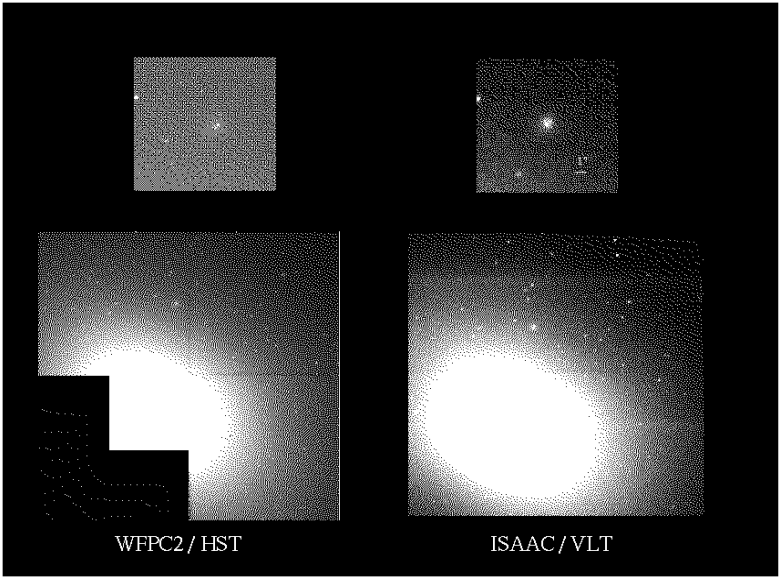,height=10cm,width=8cm}
\caption{NGC 4365 observed in the V band with the WFPC2 on HST, and in
the K band with ISAAC on the VLT. The field of views are similar; the
resolution is $\sim 0.1"$ in the HST images, $\sim 0.4"$ in the K
images; the depths are comparable. This illustrates that optical and
near-infrared imaging are becoming more and more similar for purposes of
studying globular cluster systems (images provided by T.H.~Puzia).}
\end{figure}

\subsection{Near-infrared photometry}

\begin{figure}[h]
\centerline{\psfig{figure=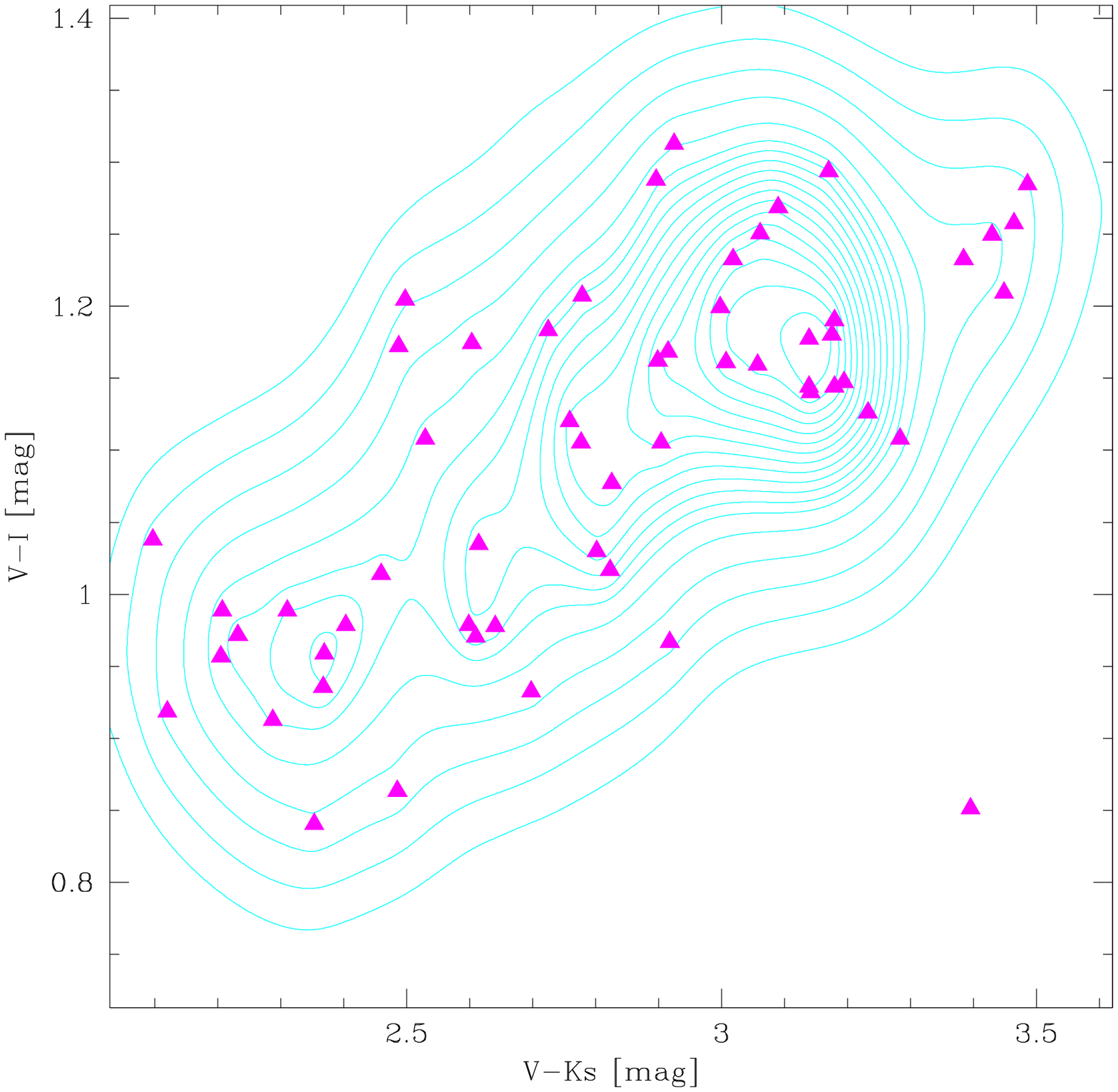,height=8cm,width=8cm
,bbllx=8mm,bblly=63mm,bburx=205mm,bbury=245mm}}
\caption{Globular cluster colors shown in the (V$-$I)--(V$-$K) plane. 
For this plot, data for NGC 3115 from ISAAC on the VLT and WFPC2 onboard HST
were combined to allow a better separation of the two sub-populations
(taken from Puzia et al.~2000, in preparation). Density contours show
the color peaks for two main sub-populations.}
\end{figure}

Since the introduction of 1024$\times$1024 pixel arrays in the near
infrared a couple of years ago, imaging at wavelength from 1.2$\mu$m to 
2.5$\mu$m became competitive in terms of depths and field size with
optical imaging (see Figure 2).

Historically, the first near-infrared measurements of extragalactic globular
clusters were carried out in M31 (Frogel, Persson \& Cohen 1980) and the
Large Magellanic Cloud (Persson et al.~1983). 
Why extend the wavelength range to the near infrared? For old globular
clusters, $V-K$ is a measure of the temperature of the red giant branch 
that is directly dependent on metallicity but hardly on age. $V-K$ is even 
more sensitive to metallicity than the Washington $C-T_1$ index. The
combination of optical and near-infrared colors is therefore superior to
optical imaging alone, both for deriving metallicities, and for
a clean separation of cluster sub-populations (see Figure 3). 
It is also used to detect potential sub-populations were optical colors 
failed to reveal any.

In young populations, $V-K$ is most sensitive to the asymptotic giant
branch which dominates the light of populations that are 0.2 to 1 Gyr old.
The combination of optical and near infrared colors can be used to
derive ages (and metallicities) of these populations (e.g.~Maraston, 
Kissler-Patig, \& Brodie 2000). 

The disadvantages of complementing optical with near-infrared colors is 
the need for a second instrument (usually a second observing run) in addition 
to the optical one. Near-infrared observations will continue fighting
against the high sky background in addition to the background light of
the galaxy which requires blank sky observations. Overall, obtaining 
near-infrared data is still very time consuming when compared to optical
studies. For example, a deep $K$ image of a galaxy will require a full
night of observations. Currently both depth and field size do not allow the near
infrared to fully replace optical colors for the study of morphological
properties or the globular cluster luminosity function. But this might
happen in the future whth the NGST.

The immediate future looks bright, with a number of ``wide-field''
imagers being available, such as ISAAC on the VLT, SOFI on the 3.5m NTT,
the Omega systems on the 3.5m Calar Alto, etc.. and 2k $\times$ 2k 
infrared arrays coming soon. The ideal future
instrument would have a dichroic which would allow to observe
simultaneously in the near-infrared and the optical.

\begin{figure}[t]
\psfig{figure=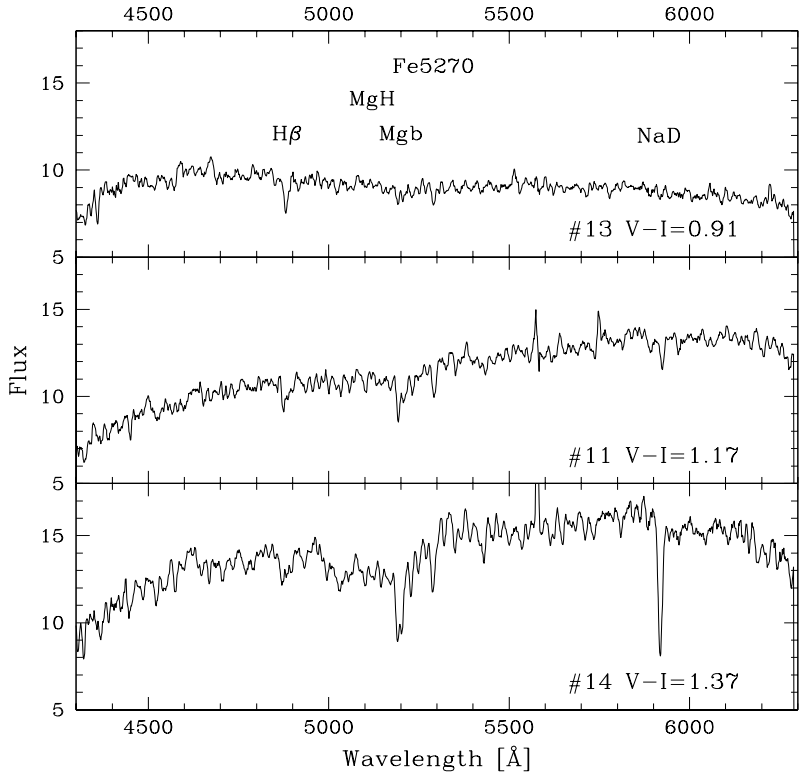,height=10cm,width=10cm
,bbllx=8mm,bblly=57mm,bburx=185mm,bbury=245mm}
\caption{Three representative spectra of globular clusters are shown, 
ranging from blue, 
over red, to very red color. While the H$\beta$ gets slightly weaker from the
blue to the red object, the metal lines (Mg, Fe, Na) become much
stronger (taken from Kissler-Patig et al.~1998a).
}
\end {figure}

\subsection{Multi-object spectroscopy}

Spectroscopy is the only way to unambiguously associate a globular cluster
with its host galaxy by matching their radial velocities. Also, it is
the most precise way to determine the metallicity of single objects, and
the only way to determine individual ages. Obviously, it
is also the only way to get radial velocities.
Ideally, one would like a spectrum of each globular cluster identified
from imaging.

In practice, good spectra are still hard to obtain. Early attempts
with 4m-class telescopes succeeded in obtaining radial velocities, but mostly
failed to determine reliable chemical abundances (see Sect.~6).
With the arrival of 10m-class telescopes, it became feasible to obtain
spectra with high enough signal-to-noise to derive chemical abundances
(Kissler-Patig et al.~1998a, Cohen, Blakeslee \& Ryzhov 1998).
Such studies are still limited to relatively bright objects ($V<23$) and
remain time consuming ($\sim$ 3h integration time for low-resolution
spectroscopy). Figure 4 shows a few examples of globular clusters in NGC
1399. High-resolution spectroscopy in order to measure internal
velocity dispersions of individual clusters is still out of reach for
old clusters, and was only carried out for two nearby super star
clusters (Ho \& Fillipenko 1996a,b), in addition to several clusters
within the Local Group. Even low-resolution spectroscopy
is currently still limited to follow-ups on photometric studies,
targeting a number of selected, representative clusters, rather than
building up own spectroscopic samples.

Current problems are the low signal-to-noise, even with 10m-telescopes,
that prohibit very accurate age or metallicity determinations for
individual clusters. The multiplexity of the existing instruments 
(FORS1 \& 2 on the VLT, LRIS on Keck) is low and only allows to spectroscopy 
a limited number of selected targets. 
Finally, the absorption indices that
are being measured on the spectra in order to determine the various
element abundances are not optimally defined. These indices lie in
the region 3800\AA\ to 6000\AA\ and were designed for spectra with 8\AA\
to 9\AA\ resolution. They often include a number of absorption lines in the
bandpass (or pseudo-continuum) other than the element to be measured. This
introduces an additional dependence e.g.~of the Balmer indices on metallicity,
etc... Using a slightly higher resolution might help defining better
indices. 

The immediate future of spectroscopy are instruments such as VIMOS on
the VLT or DEIMOS on Keck that will allow a multiplexity of 100 to 150.
These will allow to increase the exposure times and slightly the spectral 
resolution to solve a number of the problems mentioned above. These
will also allow to obtain several hundred radial velocities of globular
clusters around a given galaxy in a single night, improving
significantly the potential of kinematical studies (see Sect.~6).


\section{Globular clusters in various galaxy types, and what we learned from
them}

In this section, we present some properties of globular cluster systems and 
of young clusters in ellipticals, spirals and mergers. In the last
section we mentioned what are the properties measured in globular cluster
systems: The metallicity distribution can be obtained from photometry
(colors) or spectroscopy (absorption line indices). The luminosity
function of the clusters is computed from the measured magnitudes folded
with any incompleteness or contamination function. The total number of
clusters (and eventually number of metal-poor and metal-rich clusters) is
obtained by extrapolating the observed counts over the luminosity
function and eventually applying any geometrical completeness for the
regions that are not covered. For the latter, one uses also knowledge
about the spatial distribution (position angle, ellipticity) and radial 
density profile of the globular cluster system. For young star clusters,
the color distribution no longer reflects the metallicity distribution,
but a mix of ages and metallicities. More complex comparisons with
population synthesis models and/or spectroscopy are needed to
disentangle the two quantities. Of interest for young clusters are also
the mass distribution (derived from the luminosity function) that helps
to understand how many of the newly formed clusters will indeed evolve
into massive ``globular'' clusters.

\subsection{Globular clusters in early-type galaxies}

Early-type galaxies have the best studied globular cluster systems.
Spirals have the two systems studied in most details (i.e.~the
Milky Way and M31) due to our biased location in space, but a far larger sample
now exists for early-type galaxies.

Despite looking remarkably similar in many respects (e.g.~globular cluster
luminosity function), globular cluster
systems in early-type galaxies also show a large scatter in a number
of properties. For example, the number of globular clusters normalized
to the galaxy light (specific frequency, see Harris \& van den Bergh
1981) appears to scatter by a factor of several, mainly driven by the very 
high specific frequencies of central giant ellipticals (and recently
also observed in faint dwarf galaxies, see Durrel et al.~1996, Miller et 
al.~1998). 
Furthermore, the radial density profiles are very extended for large galaxies,
while following the galaxy light in the case of intermediate ellipticals
(e.g.~Kissler-Patig 1997a).

In the early 90's, Zepf \& Ashman (1993) discovered the presence of
globular cluster sub-populations in several early-type galaxies. We will
come back to the origin of the sub-populations in Sect.~5. Here, we will
discuss the implications of sub-populations on our understanding of the 
globular cluster system properties.

\begin{figure}[h]
\centerline{
\psfig{figure=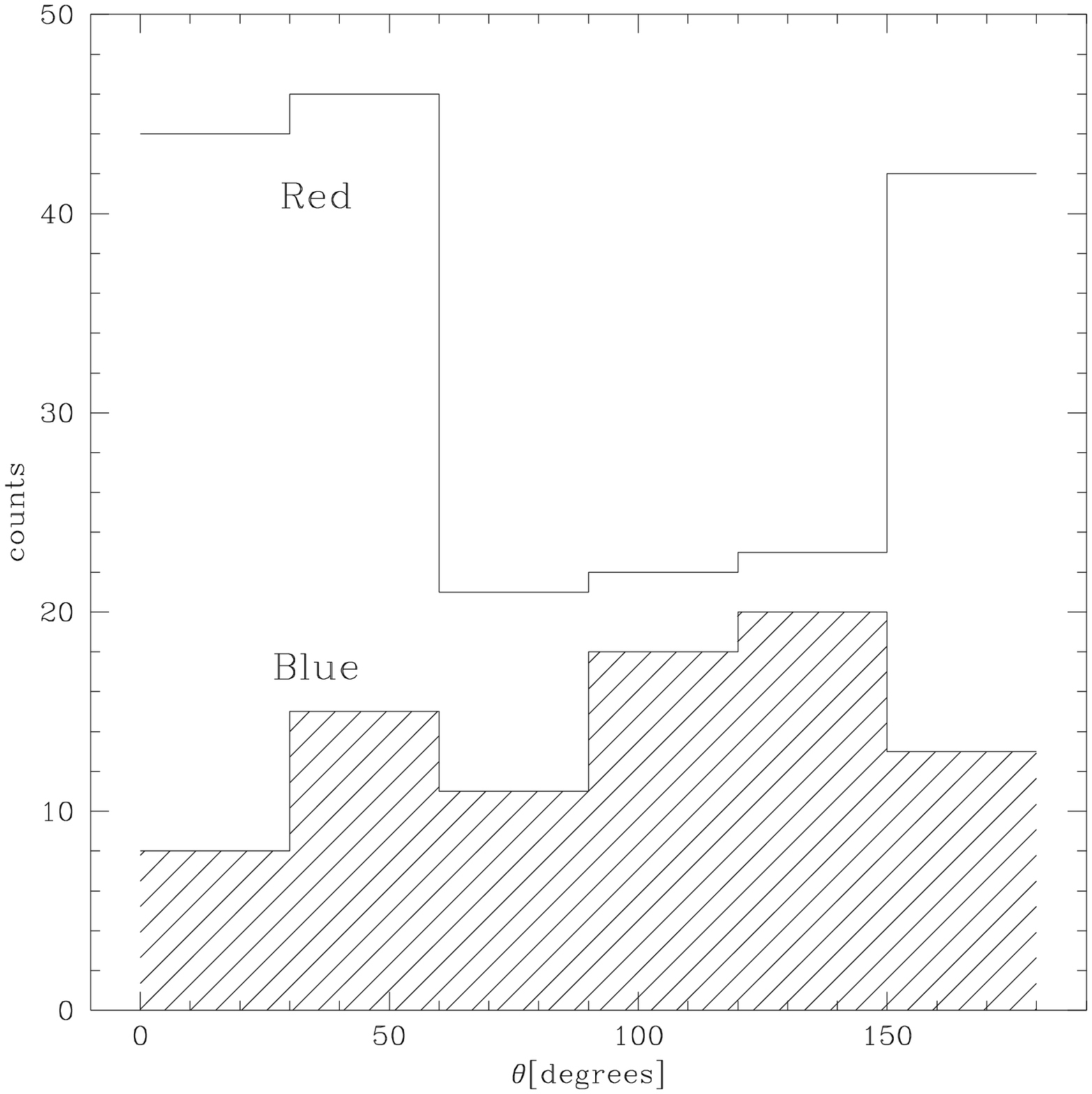,height=6cm,width=6cm
,bbllx=8mm,bblly=63mm,bburx=205mm,bbury=245mm}
\psfig{figure=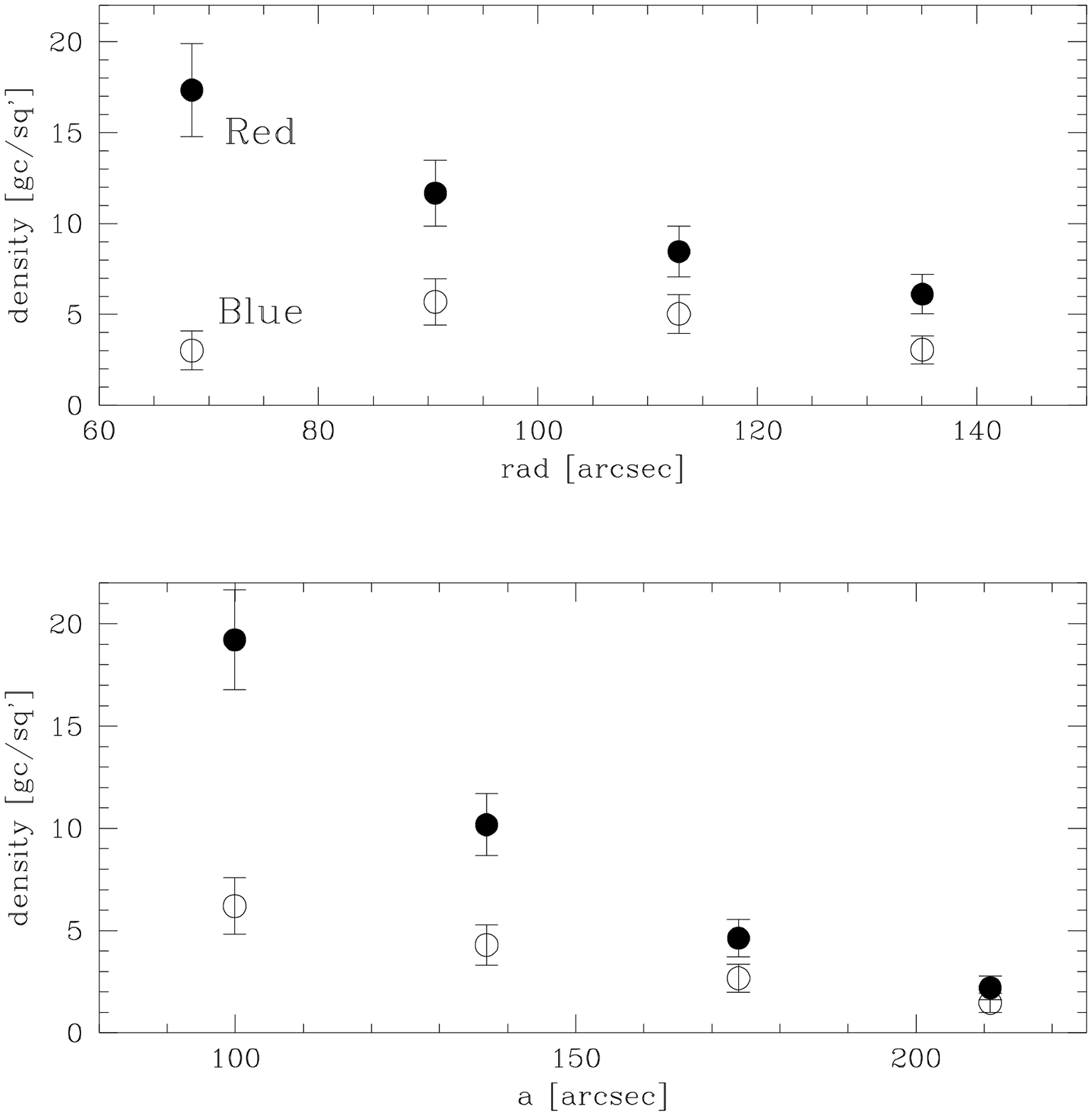,height=6cm,width=6cm
,bbllx=8mm,bblly=57mm,bburx=205mm,bbury=245mm}
}
\caption{
Left panel: The angular distribution of halo and bulge globular clusters
around NGC 1380 in 30 degree sectors, after a point symmetry around the center
of the galaxy. Note that the blue objects are spherically distributed,
while the red objects have an elliptical distribution that peaks at the
position angle of the diffuse stellar light.
Right panel: Surface density profiles of red and blue globular clusters
around NGC 1380, plotted once against the radius in arcseconds (upper panel)
and once against the semi-major axis (lower panel). Note how the blue
objects have a much flatter density profile than the red ones, which are
concentrated towards the center and follow a similar density profile as
the stellar light. Both plots are taken from Kissler-Patig et al.~(1997).
}
\end{figure}

Until the early 90's, properties were derived for the whole globular
cluster system. Since then, it became clear that many
properties need to take into account the existence of (at least two) different
sub-populations, in order to be explained. Probably the first work to
show this most clearly was the presentation of the properties of blue
and red clusters in NGC 4472 by Geisler et al.~(1996). Taking into
account the existence and different spatial distribution of blue and red
clusters, they explained two properties of whole systems at once. First, the
color gradient observed in several systems could be explained by a
varying ratio of blue to red clusters with radius (without any gradient
in the individual sub-populations). Second, the mean color of the
systems was previously thought to be systematically bluer than the diffuse
galaxy light. It turns out that the color of the red sub-population
matches the color of the galaxy, while it is the presence of the blue
``halo'' population that makes the color of the whole globular cluster
system appear bluish. 

It has not yet been demonstrated that the scatter in the specific
frequency and in the slopes of the radial density profiles also
originate from different mixes of blue to red sub-populations, but this
could be the case. The few studies that investigated separately
the morphological properties of blue and red clusters (Geisler et
al.~1996, Kissler-Patig et al.~1997, Lee et al.~1998, Kundu \& Whitmore
1998) found the metal-poor (blue) population to be more spherically
distributed and extended than the metal-rich population that
has a steeper density profile, tends to be more flattened and appears to
follow the diffuse stellar light of the galaxy in ellipticity and
position angle (cf.~Fig.~5). Thus, a larger fraction of blue clusters
in a galaxy would mimic a flatter density profile of the whole globular
cluster system. 

Furthermore, the specific frequency of the blue clusters (when related
to the blue light) appears to be very high ($>30$ see Harris 2000).
This, by the way, could be explained if the latter came from small fragments
similar to the dwarf ellipticals observed today, that also show high
specific frequency values (although not as high, but in the range 10 to 20). 
Thus, an overabundance of blue clusters would
imply a high specific frequency. Incidentally, the shallow density
profiles are found in the galaxies with the highest specific frequencies
(see Kissler-Patig 1997a). We can therefore speculate that the properties
of the entire globular cluster systems of these massive (often central)
giant ellipticals can be explained by a large overabundance of
metal-poor globular clusters originating from small fragments. The
scatter in the globular cluster system properties among ellipticals
could then (at least partly) be explained by a varying fraction of
metal-poor ``halo'' and metal-rich ``bulge'' globular
clusters.

Observationally, this could be verified by determining the number ratios
of metal-rich and metal-poor globular clusters in a sample of galaxies
showing different globular cluster radial density profiles and specific
frequencies. The number of studies investigating the
properties of metal-poor and metal-rich populations needs to increase in
order to confirm the general properties of these two groups.
We end with a word of caution: the existence of such sub-populations has
been observed in only $\sim 50$\% of all early-type galaxies studied
(e.g.~Gebhardt \& Kissler-Patig 1999), and still remains to be
demonstrated in all cases. Furthermore, the exact formation process of
these sub-populations is still unclear (see Sect.~5).

\subsection{Globular clusters in late-type galaxies}

The study of globular cluster systems of late-type spirals started with
the work of Shapley (1918) on the Milky Way system. Despite a head-start
of nearly 40 years compared to studies in early-type galaxies, the
number of studied systems in spirals lags far behind the one in
ellipticals. This is mainly due to the observational difficulties:
globular clusters in spirals are difficult to identify on the
inhomogeneous background of disks. Furthermore, internal extinction in
the spiral galaxies make detection and completeness estimations
difficult, and photometry further suffers from confusion by reddened H{\tt
II} regions, open clusters or star forming regions.

The best studied cases (Milky Way and M31) show sub-populations
(e.g.~Morgan 1959, Kinman 1959, Zinn 1985; Ashman \& Bird 1993,
Barmby et al.~1999) associated in our Galaxy with the
halo and the bulge/thick disk (Minniti 1995, C\^ot\'e 1999). Beyond the
local group, spectroscopy is needed to separate potential
sub-populations. Both abundances and kinematics are needed, while colors
suffer too much from reddening to serve as useful metallicity
tracers. Spectroscopic studies have been rare in the past, but are now
becoming feasible (see Sect.~3.3 and 4.1). For example, a recent study
of M81 allowed to identify a potential thick disk population beside
halo and bulge populations (see Schroder et al.~2000 and references therein).

Some of the open questions are whether all spirals host halo and bulge
clusters, and whether one or both populations are related to the
metal-poor and metal-rich populations in early-type galaxies. 
The number of globular clusters as traced by the specific frequency appears
roughly constant in spirals of all types independently of the presence
of a bulge and/or thick disk (e.g.~Kissler-Patig et al.~1999a). This
would mean that spirals are dominated by metal-poor populations, with
their globular cluster systems only marginally affected by the presence
of a bulge/thick disk. If metal-poor globular clusters indeed formed in
pre-galactic fragments, then one would expect the metal-poor populations
in spirals and ellipticals to be the same. We know that the globular
cluster luminosity functions are extremely similar, but the metallicity
distributions and other properties remain to be derived and compared
(see Burgarella et al.~2000 for a first attempt).
Finally, a good understanding of the globular cluster systems in spirals
will also help predicting the resulting globular cluster system of a
spiral--spiral merger. Predictions can then be compared to the
properties of systems of elliptical galaxies in order to constrain this
mode of galaxy formation. 

\subsection{Star clusters in mergers and violent interactions}

After some speculations and predictions that massive star clusters
could/should form in mergers (Harris 1981, Schweizer 1987), these were finally
discovered in the early 90's (Lutz 1991, Holtzman et al.~1992). Since
then a number of studies focussed on the detection and properties of
these massive young star clusters (see Schweizer 1997, and reviews cited in
Sect.~1 for an overview).

The most intense debate around these young clusters focussed on whether or not
their properties were compatible with a formation of early-type galaxies
through spiral--spiral mergers. It was noticed early on (Harris 1981,
van den Bergh 1982) that ellipticals appeared to host more clusters than
spirals, and thus that mergers would have to produce a large number of
globular clusters. Moreover, the specific frequency of ellipticals
appeared higher than in spirals, i.e.~mergers were supposed to form
globular clusters extremely efficiently. 
In a second stage, a number of studies investigated whether or not these
newly formed clusters would resemble globular clusters, and/or would survive
as bound clusters at all.  

The above questions are still open, except maybe for the last one. The
young clusters studied to date show luminosities, sizes, and masses
(when they can be measured) that are compatible with them being bound
stellar clusters and able to survive the next several Gyr (see Schweizer 1997 
for a summary of the studies and extensive references).
Whether they will have the exact same properties as old globular
clusters in our Milky Way is still controversial. First spectroscopic
measurements found the young clusters in NGC 7252 compatible with a
normal initial mass function (IMF) (Schweizer \& Seitzer 1998), while in
NGC 1275 the young clusters show anomalies and potentially have a
flatter IMF (Brodie et al.~1998) which would compromise their evolution
into old globular clusters, as we know them from the Galaxy.

The mass distribution of these young cluster was first found to be a
power-law (e.g.~Meurer 1995), as opposed to a log-normal distribution
for old globular clusters. This result is likely to suffer from
uncertainties in the conversion of luminosities into masses, when neglecting 
the significant age spread among the young clusters (see Fritze-von 
Alvensleben 1999). However, deeper data seem to rule out the possibility
that the initial mass distribution has already the same shape as the one observed
for old clusters (see Whitmore et al.~1999, Zepf et al.~1999).
But the slope of the mass distributions could be affected during the
evolution of the system by dynamical destruction at the low-mass end.
Finally, Whitmore et al.~(1999) recently found a break in the mass
function of the young clusters of the Antennae galaxies, similar to the
characteristic mass of the old clusters further supporting similar mass
functions for young and old cluster populations (see also Sect.~7).
Overall, the young clusters might or might not resemble old Galactic
globular clusters, but some will survive as massive star clusters and
could mimic a population of metal-rich globular clusters.

The most interesting point remains the number of clusters produced in
mergers. Obviously, this will depend on the gas content (`fuel') that is
provided by the merger (e.g.~Kissler-Patig, et al.~1998b).
Most gas-rich mergers form a large number of star clusters, but few of
the latter have masses that would actually allow them to evolve into massive
globular clusters as we observe them in distant ellipticals. Harris
(2000) reviews comprehensively this issue and other problems related
with a scenario in which all metal-rich globular clusters of ellipticals
would have formed in mergers. The main problem with such a scenario is that
the high specific frequency of ellipticals should be due to metal-rich
clusters, which is usually not the case. Potential other problems,
depending on the exact enrichment history, are that large ellipticals would be
build up by a series of mergers that should probably produce an even broader
metallicity distribution than observed; and that radial metallicity gradients
might be expected to be steeper in high specific frequency ellipticals. 

In summary, mergers are the best laboratories to study younger stellar
populations and the formation of young stellar cluster, but how
important they are in the building of globular cluster systems (and
galaxies) remains
uncertain. However, a good understanding of these clusters is crucial
for the understanding of globular cluster systems in early-type
galaxies, since merger events must have played a role at some stage.


\section{Globular cluster sub-populations and their origin}

In this section we come back to the presence of multiple sub-population
of globular clusters around a number of giant galaxies. We will briefly
review the different scenarios present in the literature that could
explain the properties of such composite systems and discuss their pros
and cons.

Sub-populations of globular clusters were first identified in the Milky
Way (Morgan 1959, Kinman 1959, Zinn 1985), and associated with 
the halo (in the case of the
metal-poor population) and the ``disk'' (in the case of the metal-rich
population. The ``disk'' clusters are now better associated with the
bulge (e.g.~Minniti 1995, C\^ot\'e 1999). The presence of multiple
component populations in other giant galaxies was 
first detected by Zepf \& Ashman (1993).
Obviously the multiple sub-populations get associated with several
distinct epochs or mechanisms of star/cluster formation.

The simple scenario of a disk--disk merger explaining the presence of
two populations of globular clusters (Ashman \& Zepf 1992) found a
strong support in the community for 5-6 years, partly because of a lack
of alternatives. It was backed up by the 
discovery of newly formed, young star clusters in interacting galaxies
(Lutz et al.~1991, Holtzman et al.~1992). Only recently, other scenarios
explaining the presence of at least two distinct populations were
presented and discussed.

\subsection{The different scenarios for sub-populations}

We will make a (somewhat artificial) separation in four scenarios and
briefly outline them and their predictions. 

\vskip 3mm
{\it The merger scenario}

The fact that mergers could produce new globular clusters was mentioned
in the literature early after Toomre (1977) proposed that ellipticals
could form out of the merging of two spirals (see Harris 1981 and Schweizer 
1987).
But the first crude predictions of the spiral-spiral merger scenario go back to
Ashman \& Zepf (1992). They predicted two populations of globular
clusters in the resulting galaxy: one old, metal-poor population from the
progenitor spirals and one newly formed, young, metal-rich population.
The metal-poor population would be more extended and would have been
transfered some of the orbital angular momentum by the merger. The
metal-rich globular clusters would be more concentrated towards the
center and probably on more radial orbits.

\vskip 3mm
{\it In situ scenarios}

In situ scenarios see all globular clusters forming within the entity
that will become the final galaxy. In this scenario, globular clusters
form in the collapse of the galaxy, which happens in two distinct phases
(see Forbes et al.~1997, Harris et al.~1998, Harris et al.~1999). The
first burst produces metal-poor globular clusters and stars (similar to
Searle \& Zinn 1978) and provokes
its own end e.g.~by ionizing the gas or expelling it (e.g.~Harris et
al.~1998). The second collapse happens shortly later (1-2 Gyr) and is at
the origin of the metal-rich component. Both populations are linked with the
initial galaxy.

\vskip 3mm
{\it Accretion scenarios}

Accretion scenarios were reconsidered in detail to explain the presence
of the large populations of metal-poor globular clusters around
early-type galaxies. In these scenarios, the metal-rich clusters belong
to the seed galaxy, while the metal-poor clusters are accreted from or
with dwarf galaxies (e.g.~Richtler 1994). 
C\^ot\'e et al.~(1998) showed in extensive
simulations that the color distributions could be reproduced. Hilker (1998) 
and Hilker et al.~(1999) proposed the accretion of stellar as well as
gas-rich dwarfs that would form new globular when accreted.
In such scenarios, the metal-poor clusters
would not be related to the final galaxies but rather have properties
compatible with that of globular clusters in dwarf galaxies. 
Furthermore, this scenario is the
only one that could easily explain metal-poor cluster that are younger than 
metal-rich ones.
In a slightly differently scenario, Kissler-Patig et al.~(1999b) mentioned the
possibility that central giant ellipticals could have accreted both metal-poor
and metal-rich clusters from surrounding medium-sized galaxies. 

\vskip 3mm
{\it Pre-galactic scenarios}

Pre-galactic scenarios were proposed long ago by Peebles \& Dicke (1968),
when the Jeans mass in the early universe was similar to globular
cluster masses. Meanwhile, it was reconsidered in the frame of globular
cluster systems (Kissler-Patig 1997b, Kissler-Patig et al.~1998b,
Burgarella et al.~2000). The metal-poor globular clusters would have
formed in fragments before the assembly of the galaxy, later-on building
up the galaxy halos and feeding with gas the formation of the bulge. In
that scenario too, the metal-poor globular clusters do not have
properties dependent from the final galaxy, while the metal-rich clusters do.
Also, metal-poor clusters are older than metal-rich clusters.

\vskip 3mm
Overall, the scenarios are discussed in the literature as different but
do not differ by much. The first scenario explains the presence of the 
metal-rich population, as opposed to the last two that deal with the 
metal-poor population.  These three scenarios are mutually not exclusive.
Only in situ models connect the metal-rich and metal-poor components.
For the metal-rich clusters, the question resumes to whether they formed 
during the collapse of the bulge/spheroid, or whether they formed in a
violent interaction. Although an early, gas-rich merger event at the
origin of the bulge/spheroid would qualify for both scenarios.
In the case of metal-poor clusters, 
the difference between the last three scenarios is mostly semantics.
They differ slightly on when the clusters formed, and models two and four might
expect differences in whether or not the properties of the clusters are
related to the final galaxy.
But the bottom line is that the boarder-line between the
scenarios is not very clear. 
Explaining the building up of globular cluster systems is probably
a matter of finding the right mix of the above mechanisms, and this for
every given galaxy.

\subsection{Pros and cons of the scenarios}

The predictions of the different scenarios are fairly fuzzy, and no scenario 
makes clear, unique predictions. Nevertheless, we can present the pros and cons
to outline potential problems with any of them. 

\vskip 3mm
{\it The merger scenario}

Pros: we know that new star cluster form in mergers (e.g.~above
mentioned reviews, and see Schweizer 1997), and will populate the
metal-rich sub-population of the resulting galaxy. Note also, that the
merger scenario is the only one that predicted bimodal color
distributions rather then explaining them after fact.

Cons: we do not know {\it i)} if all early-type galaxies formed in
mergers, {\it ii)} if the star clusters formed in mergers will indeed 
evolve into globular clusters (e.g.~Brodie et al.~1998), {\it iii)} if
all mergers produce a large number of clusters (which depends on the gas
content). Furthermore, we would then expect the metal-rich populations
to be significantly younger in many galaxies (according e.g.~to the
merger histories predicted by hierarchical clustering models). There are
still problems in explaining the specific frequencies and the right mix
of blue and red clusters in early-type galaxies in the frame of the
merger scenario (e.g.~Forbes et al.~1997).

\vskip 3mm
{\it In situ scenarios}

Pros: Searle \& Zinn (1978) list the evidences for our Milky Way halo
globular clusters to have formed in fragments building up the halo. The
massive stars in such a population would quickly create a hold of the 
star/cluster formation for a Gyr or two.

Cons: if a correlation between metal-poor clusters and galaxy is
expected, the scenario would be ruled out. A clear age sequence from
metal-poor to metal-rich clusters is predicted but not yet verified.
This scenarios is not in line with hierarchical clustering models
for the formation of galaxies (Kauffmann et al.~1993, Cole et al.~1994), 
should the latter turn out to be the right model for galaxy formation.

\vskip 3mm
{\it Accretion scenarios}

Pros: dwarf galaxies are seen in great numbers around giant galaxies,
and hierarchical clustering scenarios predict even more at
early epochs. Dwarf galaxies do get accreted (e.g.~Sagittarius in our
Galaxy). We observe ``free-floating'' populations around central cluster
galaxies (e.g.~Hilker et al.~1999) and the color distributions of
globular cluster systems can be reproduced (C\^ot\'e et al.~1998).

Cons: we are missing detailed dynamical simulations  of galaxy groups
and clusters to test whether the predicted large number of dwarf
galaxies gets indeed accreted (and when). We do not know whether the
(dwarf) galaxy luminosity function was indeed as steep as required at
early times to explain the large accretion rates needed. 
Also, the model does not provide a physical explanation for
the metal-rich populations.

\vskip 3mm
{\it Pre-galactic scenarios}

Pros: similar to the above, we observe a ``free-floating'', spatially
extended populations of globular clusters around central galaxies. The
properties of the metal-poor populations do not seem to correlate with
the properties of their host galaxies (Burgarella et al.~2000). 
The metal-poor globular cluster are observed to be very old
(e.g.~Ortolani et al.~1995 for our Galaxy; Kissler-Patig et al.~1998a, 
Cohen et al.~1998, Puzia et al.~1999 for analogies in extragalactic
systems).

Cons: galaxies and galaxy halos might not have formed by the agglomeration of 
independent fragments. No physical model exists, except a broad
compatibility with hierarchical clustering models (see also 
Burgarella et al.~2000).

\vskip 3mm

Some pros and cons are listed only under one scenario but apply
obviously to others. It should be noted that these pros and cons apply to
``normal'' globular cluster systems. It has been noted that several
galaxies host very curious mixes of metal-poor and metal-rich clusters
(Gebhardt \& Kissler-Patig 1999, Harris et al.~2000) that pose
challenges to all scenarios. Fine difference will require a much more
detailed abundance analysis of the individual clusters in
sub-populations, as well as their dynamical properties and (at least
relative) ages for the different globular cluster populations. These
might allow to identify a unique prediction supporting the one or the
other formation mode, or constrain the importance of each formation mechanism.


\section{Kinematics of globular clusters}

\begin{figure}[h]
\hskip 1cm \psfig{figure=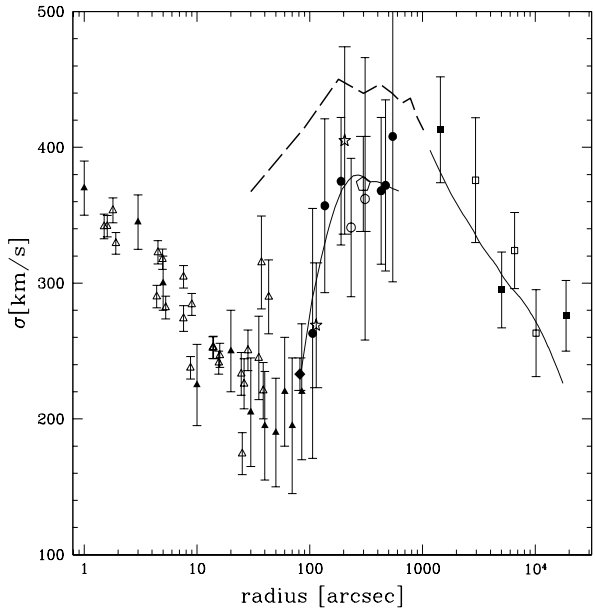,height=8cm,width=8cm
,bbllx=8mm,bblly=57mm,bburx=205mm,bbury=245mm}
\caption{Velocity dispersion as a function of radius for various
components
around NGC 1399, see Kissler-Patig et al.~(1998a) for details. The two
solid lines are fits to the velocity data of the globular clusters and
of the Fornax galaxies. The dashed line shows the X--ray temperature
converted to a velocity dispersion, the triangles are stellar
measurements.}
\end{figure}

\begin{figure}[h]
\hskip 2cm \psfig{figure=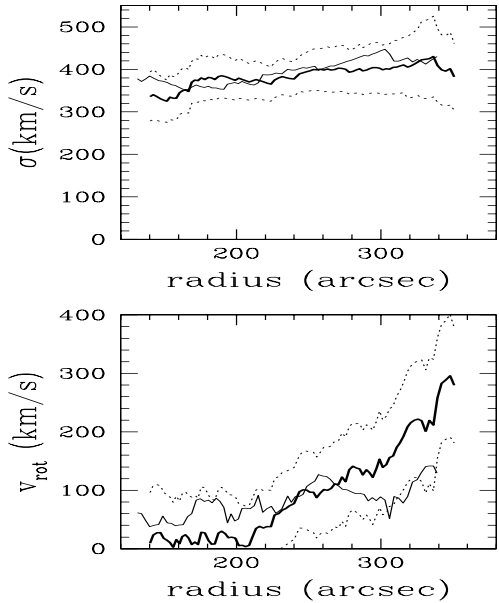,height=8cm,width=14cm
,bbllx=8mm,bblly=57mm,bburx=205mm,bbury=245mm}
\caption{Kinematics of red (thin line) and blue (thick line) globular clusters
in M87.  Projected velocity dispersion, and projected rotational velocity
as functions of radius for a fixed position angle of 120$^{\circ}$.
Dotted lines mark the 68\% confidence bands. Taken from Kissler-Patig
\& Gebhardt 1998.
}
\end{figure}

In this section, we briefly discuss recent results from kinematical
studies of extragalactic globular clusters. The required measurements
were discussed in Sect.~3. Kinematics can be used both to understand the
formation of the globular cluster systems, as well as to derive dynamics
of galaxies at large radii.

\subsection{Globular cluster system formation}

Globular cluster system kinematics are used since a long time to
constrain their formation. In the Milky Way, kinematics support the
association of the various clusters with the halo and the bulge (see
Harris 2000 and references therein). In M31, similar results were
derived (Huchra et al.~1991, Barmby et al.~1999). In M81, the
situation appears very similar again (Schroder et al.~2000). 
Beyond the Local group, radial velocities for globular clusters are somewhat
harder to obtain. Nevertheless, studies of globular cluster kinematics in
elliptical galaxies started over a decade ago (Mould et al.~1987, 1990,
Huchra \& Brodie 1987, Harris 1988, Grillmair et al.~1994).

Figure 6 illustrates one example where the kinematics of globular
clusters allowed to gain some insight into the globular cluster
system formation (from Kissler-Patig et al.~1999b). The figure
shows the velocity dispersion as a function of radius around NGC 1399, the 
central giant elliptical in Fornax.  
The velocity dispersion of the globular clusters increases with
radius, rising from a value not unlike that  for the outermost stellar
measurements at 100\arcsec ,  to values almost  twice as  high at
$\sim$ 300\arcsec . The  outer velocity dispersion measurements are in
good agreement with the temperature of  the X-ray gas and the velocity
dispersion of galaxies in the  Fornax cluster. Thus, a large fraction of  the
globular clusters  which we associate with NGC 1399 could rather be
attributed to the whole of the Fornax cluster. By association, this
would be true for the stars in the cD envelope too.
This picture strongly favors the accretion or pre-galactic scenarios for
the formation of the metal-poor clusters in this galaxy. 

As another example, Fig.~7 shows the velocity dispersion and rotational
velocity for the metal-poor and metal-rich globular clusters around
M87, the central giant elliptical in Virgo. There is some evidence that the 
rotation is confined to the metal--poor globular clusters. If, as assumed, the
last merger was mainly dissipationless (and did not form a significant
amount of metal-rich clusters), this kinematic difference
between the two sub--populations could reflect the situation 
in the {\it progenitor} galaxies of M87. These would then
be compatible (see Hernquist \& Bolte 1992) with a formation in a gas-rich
merger event (see Ashman \& Zepf 1992).

Generally, the data seem to support the view that the metal-poor
globular clusters form a hot system with some rotation, or tangentially
biased orbits. The metal-rich globular clusters have a lower velocity
dispersion in comparison, and exhibit only weak rotation, if at all
(Cohen \& Ryzhov 1997, Kissler-Patig et al.~1999b, Sharples et al.~1999,
Kissler-Patig \& Gebhardt 1999, Cohen 2000). The interpretation of these
results in the frame of the different formation scenarios presented in
Sect.~5 is unclear, since no scenario makes clear and unique predictions
for the kinematics of the clusters. Furthermore, some events unrelated
to the formation of the globular clusters can alter the dynamics:
e.g.~a late dissipationless mergers of two ellipticals could convey
angular momentum to both metal-rich and metal-poor clusters, bluring
kinematical signatures present in the past. Detailed dynamical
simulations of globular cluster accretion and galaxy mergers are
necessary in order to compare the data with scenario predictions. But
clearly, kinematics can help understanding differences in the metal-poor
and metal-rich components, exploring intra-cluster globular clusters,
and studying the formation of globular cluster systems as a whole.

\subsection{Galaxy dynamics}

Kinematical studies of globular clusters can also be used to study
galaxy dynamics. The globular clusters do only represent discrete probes in the
gravitational potential of the galaxy, as opposed to the diffuse stellar
light that can be used as a continuous probe with radius, but globular
clusters have the advantage (such as planetary nebulae) to extend
further out. Globular clusters can be measured out to several effective
radii, probing the dark halo and dynamics at large radii.

The velocity dispersion around NGC 1399, presented above, is one
example. Another example was presented by Cohen \& Ryzhov (1997) who derived 
from the velocity dispersion of
the globular clusters in M87 a mass of $3\times10^{12}M_\odot$ at 44kpc and 
a mass-to-light ratio $>30$, strongly supporting the presence of a massive
dark halo around this galaxy. With the same data, Kissler-Patig \&
Gebhardt (1998) derived a spin for M87 of $\lambda \sim 0.2$, at the
very high end of what is predicted by cosmological N-body simulations.
The authors suggested as most likely explanation for the data a major 
(dissipationless) merger as the last major event in the building of M87.

These examples illustrate what can be learned about the galaxy formation
history from kinematical studies of globular cluster system. In the
future, instruments such as VIMOS and DEIMOS will allow to get many
hundreds velocities in a single night for a given galaxy. These data
will allow to constrain even more strongly galaxy dynamics at large
radii.


\section{Globular clusters as distance indicators}

In this section, we will review the globular cluster luminosity function
(GCLF) as a distance indicator. The method is currently ``unfashionable'' in 
the literature mainly because some previous results seem to be in
contradiction with other distance indicators (e.g.~Ferrarese et al.~1999).
We will try to shade some light on the discrepancies, and show that, if
the proper corrections are applied, the GCLF competes well with other
extragalactic distance indicators. 

\subsection{The globular cluster luminosity function}

A nice overview of the method is given in Harris (2000), including some
historical remarks and a detailed description of the method. A further
review on the GCLF method was written by Whitmore (1997), who
addressed in particular the errors accompanying the method.
We will only briefly summarize the method here.

The basics of the method are to measure in a given filter (most often
$V$) the apparent magnitudes of a large number of globular clusters in
the system. The so constructed magnitude distribution, or luminosity
function, peaks at a characteristic (turn-over) magnitude. The absolute value
for this characteristic magnitude is derived from local or secondary
distance calibrators, allowing to derive a distance modulus from the observed
turn-over magnitude. Figure 8 shows a typical globular cluster luminosity
function with its clear peak (taken from Della Valle et al.~1998).

\begin{figure}[t]
\psfig{figure=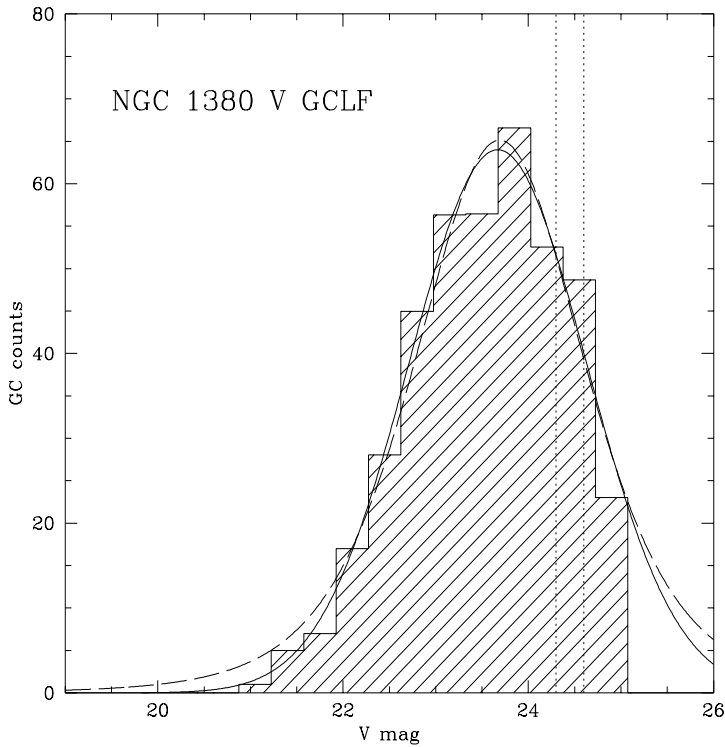,height=10cm,width=10cm
,bbllx=8mm,bblly=57mm,bburx=205mm,bbury=245mm}
\caption{Typical GCLF, observed in the $V$ band in the galaxy NGC 1380
(Della Valle et al.~1998). Over-plotted are two fits (a Gaussian
and a student t$_5$ function). The dotted lines show the 50\%
completeness limit in the $V$ filter and in the filter combination
($B,V,R$) that was used to select the globular clusters. The apparent peak
magnitude can be derived and the distance computed using the
absolute magnitude obtained from local or secondary calibrators.}
\end{figure}

The justification of the method is mainly empirical. 
Apparent turn-overs for galaxies at the same distance (e.g.~in the same
galaxy cluster) can be compared and a scatter around 0.15 mag is then
obtained, without
correcting for any external error. Similarly, a number of well observed
apparent turn-over magnitudes can be transformed into absolute ones using 
distances from other distance indicators (Cepheids where possible, or a mean of
Cepheids, surface brightness fluctuations, planetary luminosity
function, ...) and a similar small scatter is found (see Harris 2000 for a
recent compilation). Taking into account the errors in the
photometry, the fitting of the GCLF, the assumed distances, etc... this hints
at an internal dispersion of the turn-over magnitude of $<0.1$, making it a
good standard candle. From a theoretical point of view, this constancy
of the turn-over magnitude translates into a ``universal''
characteristic mass in the globular cluster mass distributions in all
galaxies. Whether this is a relict of a characteristic mass in the mass
function of the molecular clouds at the origin of the globular clusters,
or whether it was implemented during the formation process of the
globular clusters is still unclear.

The absolute turn-over magnitude lies around $V_{TO}\sim-7.5$, and the 
determination of the visual turn-over is only accurate if the peak of
the GCLF is reached by the observations. From an observational point of
view, this means that the data must reach e.g.~$V\sim 25$ to determine
distances in the Fornax or Virgo galaxy clusters (D$\sim 20$ Mpc), and
that with HST or 10m-class telescopes reaching typically $V\sim 28$, the
method could be applied as far out as 120 Mpc (including the Coma galaxy
cluster).

The observational advantages of this method over others are that 
globular clusters are brighter than other standard candles (except for 
supernovae), and do not vary, i.e.~no repeated observations are
necessary. Further, they are usually measured at large radii or in the
halo of (mostly elliptical) galaxies where reddening is not a concern.

\subsection{General problems}

A large number of distance determinations from the GCLF were only
by-products of globular cluster system studies, and often suffered from
purely practical problems of data taken for different purposes.

First, a good estimation of the background contamination is necessary to
clean the globular cluster luminosity function from the luminosity
function of background galaxies which tends to mimic a fainter turn-over
magnitude. Next, the finding incompleteness for the globular clusters
needs to be determined, in 
particular as a function of radius since the photon noise is changing 
dramatically with galactocentric radius. Proper reddening corrections
need to be applied and might differ whether one uses the ``classical''
reddening maps of Burstein \& Heiles (1984) or the newer maps from
Schlegel et al.~(1998). When necessary, proper aperture correction for
slightly extended clusters on WFPC2/HST images has to be made (e.g.~Puzia et 
al.~1999). Finally, several different ways of fitting the GCLF are used:
from fitting a histogram, over more sophisticated maximum-likelihood
fits taking into account background contamination and incompleteness.
The functions fitted vary from Gaussians to Student (t$_5$) functions,
with or without their dispersion as a free parameter in addition to the
peak value.

In addition to these, errors in the absolute calibration will be added
(see below). Furthermore,
dependences on galaxy type and environment were claimed, although the
former is probably due to the mean metallicity of globular clusters
differing in early- and late-type galaxies, while the latter was never 
demonstrated with a reliable set of data.

All the above details can introduce errors in the analysis that might sum up
to several tenths of a magnitude. The fact that distance determinations using
the GCLF are often a by-product of studies aiming at understanding globular
cluster or galaxy formation and evolution,  did not help in constructing
a very homogeneous sample in the past. The result is a very
inhomogeneous database (e.g.~Ferrarese et al.~1999) dominated by large
random scatter introduced in the analysis, as well as systematic errors
introduced by the choice of calibration and the complex nature of
globular cluster systems (see below).
Nevertheless, most of these problems were recognized and are overcome by
better methods and data in the recent GCLF distance determinations. 

\subsection{The classical way: using all globular clusters of a system}

Harris (2000, see also Kavelaars et al.~2000) outline what we will call
the classical way of measuring distances with the GCLF. This method
implies that the GCLF is measured from all globular clusters in a
system. In addition, it uses the GCLF as a ``secondary''
distance indicator, basing its calibration on distances derived by
Cepheids an other distance indicators. The method compares the peak of the 
observed GCLF with the peak of a compilation of GCLFs from mostly Virgo and
Fornax ellipticals, adopting from the literature a distance to these 
calibrators.
This allowed, among others, Harris' group to determine a distance to Coma
ellipticals and to construct the first Hubble diagram from GCLFs in
order to derive a value for H$_0$ (Harris 2000, Kavelaars et al.~2000).

In practice, an accurate GCLF turn-over is determined (see above) and calibrated
without any further corrections using M$_V(TO) = -7.33\pm0.04$ (Harris
2000) or M$_V(TO) = -7.26\pm0.06$ using Virgo alone (Kavelaars et al.~2000).

The advantages of this approach are the following. Using all globular clusters
(instead of a limited sub-population) often avoids problems with small number
statistics. This is also the idea behind using Virgo GCLFs instead of
the spars Milky Way GCLF as calibrators. The Virgo GCLFs, derived from giant 
elliptical galaxies rich in globular clusters, are well sampled and do not 
suffer from small number statistics. Further, since most newly derived
GCLFs come from cluster ellipticals, one might be more confident
to calibrate these using Virgo (i.e.~cluster) ellipticals, in order to avoid 
any potential dependence on galaxy type and/or environment. 

However, the method has a number of caveats. The main one is that giant
ellipticals are known to have globular cluster sub-populations with
different ages and metallicities. This automatically implies that the
different sub-populations around a given galaxy will have different
turn-over magnitudes. By using the whole globular cluster systems, one
is using {\it a mix} of turn-over magnitudes. One could in principal try
to correct e.g.~for a mean metallicity (as proposed by Ashman, Conti \&
Zepf 1995), but this correction depends on the population synthesis
model adopted (see Puzia et al.~1999) and implies that the mix of
metal-poor to metal-rich globular clusters is known. This mix does not only vary
from galaxy to galaxy (e.g.~Gebhardt \& Kissler-Patig 1999), but also
varies with galactocentric radius (e.g.~Geisler et al.~1996, Kissler-Patig 
et al.~1997). It results in a displacement of the turn-over peak and the
broadening of the observed GCLF of the whole system.
The Virgo ellipticals are therefore only valid calibrators for other
giant ellipticals with a similar ratio of metal-poor to metal-rich
globular clusters {\it and} for which the observations cover similar
radii. This is potentially a problem when comparing ground-based
(wide-field) studies with HST studies focusing on the inner regions of
a galaxy. Or when comparing nearby galaxies where the center is well
sampled to very distant galaxies for which mostly halo globulars are
observed. In the worse case, ignoring the presence of different
sub-populations and comparing very different galaxies in this respect,
can introduce errors a several tenths of magnitudes.

Another caveat of the classical way, is that relative distances to Virgo
can be derived, but absolute magnitudes (and e.g.~values of H$_0$) will
still dependent on other methods such as Cepheids, surface brightness
fluctuations (SBF), Planetary Nebulae luminosity functions (PNLF), and
tip of red-giant branches (TRGB), i.e.~the method will never overcome
these other methods in accuracy and carry along any of their potential 
systematic errors.

\subsection{The alternative way: using metal-poor globular clusters only}

As an alternative to the classical way, one can focus on the metal-poor
clusters only. The idea is to isolate the metal-poor globular clusters
of a system and to determine their GCLF. As a calibrator, one can use
the GCLF of the metal-poor globular clusters in the Milky Way, which
avoids any assumption on the distance of the LMC and will be independent
of any other extragalactic distance indicator. For the Milky Way GCLF,
the idea is to re-derive an absolute distance to each individual cluster,
resulting in individual absolute magnitudes and allowing to derive an absolute
luminosity function. Individual distances to the clusters are derived using the
known apparent magnitudes of their horizontal branches and a relation
between the absolute magnitude of the horizontal branch and the
metallicity (e.g.~Gratton et al.~1997). The latter is based on {\tt
HIPPARCOS} distances to sub-dwarfs fitted to the lower main sequence of
chosen clusters. This methods follows a completely different path
than methods based at some stage on Cepheids. In particular, the method
is completely independently from the distance to the LMC.

In practice, an accurate GCLF turn-over (see above) for the metal-poor clusters
in the target galaxy
is derived and calibrated, without any further corrections, using M$_V(TO)=
-7.62\pm0.06$ derived from the metal-poor clusters of the Milky Way (see
Della Valle et al.~1998, Drenkhahn \& Richtler 1999; note that the error
is statistical only and does not include any potential systematic error
associated with the distance to Galatic globular clusters, currently
under debate).

The advantages of this method are the following. This method takes into
account the known sub-structures of globular cluster systems. Using the
metal-poor globular clusters is motivated by several facts. First, they
appear to have a true universal origin (see Burgarella et al.~2000), and
their properties seem to be relatively independent of galaxy type,
environment, size and metallicity. Thus, to first order they can be used
in all galaxies without applying any corrections. In addition, the Milky Way is
justified as calibrator even for GCLFs derived from elliptical galaxies.
Further, they appear to be ``halo'' objects, i.e.~little affected by
destruction processes that might have shaped the GCLF in the inner few
kpc of large galaxies, or that affect objects on radial orbits. They
will certainly form a much more homogeneous populations than the total
globular cluster system (see previous sections).
Using the Milky Way as calibrator allows this method to be completely
independent on other distance indicators and to check independently
derived distances and value of H$_0$.

The method is not free from disadvantages. First, selecting metal-poor
globular clusters requires better data than are currently used in most
GCLF studies, implying more complicated and time-consuming observations. 
Second, even with excellent data  a perfect separation of metal-poor and
metal-rich clusters will not be
possible and the sample will be somewhat contaminated by metal-rich
clusters. Worse, the sample size will be roughly halved (for a typical
ratio of blue to red clusters around one). This
might mean that in some galaxies less than hundred clusters will be
available to construct the luminosity function, inducing error $>0.1$ on
the peak determination due to sample size alone.
Finally, the same concerns applies as for the whole sample: how
universal is the GCLF peak of metal-poor globular clusters? This remains
to be checked, but since variations of the order of $<0.1$ seem to be
the rule for whole samples, there is no reason to expect a much larger
scatter for metal-poor clusters alone.

\subsection{A few examples, comparisons, and the value of H$_0$ from the
GCLF method}

Two examples of distance determinations from metal-poor clusters were
given in Della Valle et al.~(1998), and Puzia et al.~(1999).

The first study derived a distance modulus for NGC 1380 in the Fornax
cluster of $(m-M)=31.35\pm0.09$ (not including a potential systematic error of
up to 0.2). In this case, the GCLF of the metal-poor and the metal-rich
clusters peaked at the same value, i.e.~the higher metallicity was
compensated by a younger age (few Gyr) of the red globular cluster
population, so that it would not make a difference whether one uses the
metal-poor clusters alone or the whole system. As a comparison, values
derived from Cepheids and a mean of Cepheids/SBF/PNLF to Fornax are
$(m-M)=31.54\pm0.14$ (Ferrarese et al.~1999)  and $(m-M)=31.30\pm0.04$
(from Kavelaars et al.~2000).

In the case of NGC 4472 in the Virgo galaxy cluster, 
Puzia et al.~(1999) derived turn-overs from the
metal-poor and metal-rich clusters of $23.67\pm0.09$ and $24.13\pm0.11$
respectively. Using the metal-poor clusters alone, their derived
distance is then $(m-M)=30.99\pm0.11$. This compares with the Cepheid
distance to Virgo from 6 galaxies of $(m-M)=31.01\pm0.07$ and to the
mean of Cepheids/SBF/TRGB/PNLF of $(m-M)=30.99\pm0.04$ (from Kavelaars
et al.~2000). Both cases show clearly the excellent agreement of the
GCLF method with other popular methods, {\it despite the completely
different and independent calibrators used}. The accuracy of the GCLF method
will always be limited by the sample size and lies around $\sim 0.1$.

A nice example of the ``classical way'' is the recent determination of
the distance to Coma. At the distance of $\sim 100$ Mpc the separation
of metal-poor and metal-rich globular clusters is barely feasible
anymore, and using the full globular cluster systems is necessary.
Kavelaars et al.~(2000) derived turn-over values of
M$_V(TO)=27.82\pm0.13$ and M$_V(TO)=27.72\pm0.20$ for the two galaxies
NGC 4874 and IC 4051 in Coma, respectively. Using Virgo ellipticals as
calibrators and assuming a distance to Virgo of $(m-M)=30.99\pm0.04$,
they derive a distance to Coma of $102\pm6$ Mpc. Adding several
turn-over values for distant galaxies (taken from Lauer et al.~1998),
they construct a Hubble diagram for the GCLF technique and derive a
Hubble constant of H$_0=69\pm9$ km s$^{-1}$ Mpc$^{-1}$. This example
demonstrates the reach in distance of the method.

\subsection{The Future of the method}

In summary, we think that the method is mature now and that most errors
in the analysis can be avoided, as well as good choices for the
calibration made. In the future, with HST and 10m-class telescope data, a
number of determinations in the 100 Mpc range will emerge, and
eventually, using metal-poor globular clusters only, this will give us a
grasp on the distance scale completely independent from distances based
at some stage on the LMC distance or Cepheids.


\section{Conclusions}

All the previous section should have made clear that globular clusters
can be used for a very wide variety of studies. They can constrain the
star formation history of galaxies, in particular on the two or more
distinct epochs of star formation in early-type galaxies. They can help
explaining the building up of spiral galaxies, and the star formation in
violent interactions. They can be useful to study galaxy dynamics at
large galactocentric radii. And finally, they provide an accurate
distance indicator, independent of Cepheids and the distance to the LMC.
This makes the study of globular cluster systems one of the most
versatile fields in astronomy.

Extragalactic globular cluster research experienced a boom in the early
90s with the first generation of reliable CCDs, and the first imaging
from space. We can expect a continuation of the improvement of optical
imaging, but more important, the field will benefit from the advancement
in near-infrared imaging, and most of all, of the upcoming multi-object
spectrographs on 10m telescopes. The next little revolution in this
subject will come with the determination of hundreds of globular
cluster abundances around a large number of galaxies. The next 5 years
will be an exiting time.

\vspace{0.7cm}
\noindent
{\large{\bf Acknowledgments}}

First of all, I would like to thank the Astronomische Gesellschaft for awarding
me the Ludwig-Bierman Price. I feel extremely honored and proud. For his
constant support, I would like to thank Tom Richtler,
who introduced me to the fascinating subject of globular clusters. For
the most recent years, I would like to thank Jean Brodie for her collaboration
and for giving me the first
opportunity to use a 10m telescope to satisfy my curiosity. I am grateful
to my current collaborators Thomas Puzia, Claudia Maraston, Daniel
Thomas, Denis Burgarella, Veronique Buat, Sandra Chapelon, Michael
Hilker, Dante Minniti, Paul Goudfrooij, Linda Schroder and many
others, for helping me to keep up the flame. As usual, I would be lost without
Karl Gebhardt's codes and sharp ideas. I am grateful to Klaas de Boer
and Simona Zaggia for comments on various points. And last but not least, I 
am very thankful to Steve Zepf for a critical reading of the manuscript.

\vspace{0.7cm}
\noindent
{\large{\bf References}}
{\small

\ref
Ashman K.M., \& Bird C.M. 1993, AJ 106, 2281

\ref
Ashman K.M., Conti A., Zepf S.E. 1995, AJ, 110, 1164

\ref
Ashman K.M., \& Zepf S.E. 1998, ``Globular Cluster Systems'', Cambridge
University Press 

\ref 
Barmby P., Huchra J.P., Brodie J.P. et al.~2000, AJ in press

\ref
Brodie J.P., Schroder L.L., Huchra J.P. et al.~1998, AJ 116, 691

\ref
Bunker A.J., van Breugel W.J.M. 1999, ``The Hy-Redshift Universe: Galaxy
Formation and Evolution at High Redshift '', ASP Conf. Ser.

\ref 
Burgarella D., Kissler-Patig M., \& Buat V. 2000, A\&A submitted

\ref
Burstein D., \& Heiles C. 1984, ApJS 54, 33

\ref
Della Valle M., Kissler-Patig M., Danziger J., \& Storm J. 1998, MNRAS
299, 267

\ref
Durrel P.R., Harris W.E., Geisler D., \& Pudritz R.E. 1996, AJ 112, 972

\ref
Cohen J.G. 2000, AJ in press

\ref 
Cohen J.G., \& Ryzhov A. 1997, ApJ 486, 230

\ref
Cohen J.G., Blakeslee J.P., \& Ryzhov A. 1998 ApJ, 496, 808

\ref
Cole S., Arag\'on-Salamanca A., Frenk C.S., et al.~1994, MNRAS 271, 781

\ref 
Combes F., Mamon G.A., \& Charmandaris V. 1999,  ``Dynamics of Galaxies: 
from the Early Universe to the Present'' ASP Conf. Ser., Vol.197

\ref
Couture J., Harris W.E., \& Allwright J.W.B. 1991, ApJ 372, 97

\ref 
C\^ot\'e P., Marzke R.O., \& West M.J. 1998, ApJ 501, 554

\ref
C\^ot\'e P. 1999, AJ 118, 406

\ref
Drenkhahn  G., \& Richtler T. 1999, A\&A 349, 877

\ref
Ferrarese L., Ford H.C., Huchra J.P., et al.~1999, ApJS in press

\ref 
Forbes D.A., Brodie J.P., \& Grillmair C.J. 1997, AJ 113, 1652

\ref
Fritze-von Alvensleben U. 1999, A\&A 342, L25

\ref
Frogel J.A., Persson S.E. \& Cohen J.G. 1980, ApJ 240, 785

\ref
Gebhardt K., \& Kissler-Patig M. 1999, AJ 118, 1526

\ref 
Geha M.C., Holtzman J.A., Mould J.R., et al.~1998, AJ 115, 1045

\ref 
Geisler D., \& Forte J.C. 1990, ApJ 350, L5

\ref 
Geisler D., Lee M. G.,\& Kim E. 1996, AJ 111, 1529

\ref
Gratton R.G., Fusi Pecci F., Carretta E., et al.~1997, ApJ 491, 749

\ref
Grillmair C.J., Freeman K.C., Bicknell G.V., et al.~1994, ApJ 422, L9

\ref
Harris G.L.H., Harris W.E., \& Poole G.B. 1999, AJ 117, 855

\ref 
Harris H.C. 1988, in ``Globular Cluster Systems in Galaxies'', IAU
Symp.126, eds. J.E.Grindlay \& A.G.D.Philip, Dodrecht:Kluwer, p.205

\ref 
Harris W.E. 1981, ApJ 251, 497

\ref
Harris W.E. 1991, ARA\&A 29, 543

\ref 
Harris W.E. 2000, ``Globular Cluster Systems'', Lectures for the 1998
Saas-Fee Advanced School on Star Clusters, Springer

\ref
Harris W.E, \& van den Bergh S. 1981, AJ 86, 1627

\ref 
Harris W.E., Harris G.L.H., \& McLaughlin D.E. 1998, AJ 115, 1801

\ref 
Harris W.E., Kavelaars J.J., Hanes D.A., et al.~2000, ApJ 533, in press

\ref
Hernquist, L. \& Bolte, M. 1992, in ``The Globular Cluster Galaxy
Connection'', ASP Conf. Ser., Vol. 48, eds. G.Smith, J.P.Brodie, p.788

\ref
Hilker M. 1998, PhD thesis, Sternwarte Bonn

\ref
Hilker M., Infante P., \& Richtler T. 1999, A\&AS 138, 55 

\ref 
Ho L.C., \& Filipenko A.V. 1996a, ApJ 466, L83

\ref 
Ho L.C., \& Filipenko A.V. 1996b, ApJ 472, 600

\ref
Holtzman J.A., Faber S.M., Shaya E.J., et al.~1992, AJ 103, 691

\ref
Huchra J.P., \& Brodie J.P. 1987, AJ 93, 779

\ref
Huchra J.P., Kent S.M., \& Brodie J.P. 1991, ApJ 370, 495

\ref
Kauffmann G., White S.D.M., \& Guiderdoni B. 1993, MNRAS 264, 201

\ref
Kavelaars J.J., Harris W.E., Hanes D.A., et al.~2000, ApJ in press

\ref 
Kinman T.D. 1959, MNRAS 119, 538

\ref 
Kissler-Patig M. 1997a, A\&A 319, 83

\ref 
Kissler-Patig M. 1997b, PhD thesis, Sternwarte Bonn

\ref 
Kissler-Patig M., Richtler T., Storm J., Della Valle M 1997, A\&A 327, 503

\ref
Kissler-Patig M., \& Gebhardt K. 1998, AJ 116, 2237

\ref
Kissler-Patig M., Brodie J.P., Schroder L.L., et al.~1998a, AJ 115, 105

\ref
Kissler-Patig M., Forbes D.A., Minniti D. 1998b, MNRAS 298, 1123

\ref
Kissler-Patig M., Ashman K.M., Zepf S.E., \& Freeman K.C. 1999a, AJ 118, 197

\ref 
Kissler-Patig M., Grillmair C.J., Meylan G., et al.~1999b, AJ 117, 1206

\ref 
Maraston C., Kissler-Patig M., \& Brodie J.P. 2000, in preparation

\ref 
Kundu A., \& Whitmore B.C. 1998, AJ 116, 2841

\ref 
Kundu A., Whitmore B.C., Sparks W.B, et al.~1999, ApJ 513, 733

\ref
Larsen S.S., \& Richtler T. 1999, A\&A 345, 59

\ref
Lauer T.R., Tonry J.R., Postman M., et al.~1998, ApJ 499, 577

\ref
Lee M.G., Kim E., \&  Geisler D. 1998, AJ 115, 947

\ref 
Lutz D., A\&A 245, 31

\ref 
Mazure A., Le Fevre O., \& Lebrun V. 1999, "Clustering at High Redshift", 
Les Rencontres Internationales de l'IGRAP, ASP Conf. Ser.

\ref
Meurer G.R. 1995, Nature 375, 742

\ref 
Miller B.W., Lotz J.M., Ferguson H.C. et al.~1998, ApJ 508, L133

\ref
Minniti D. 1995, AJ 109, 1663

\ref
Morgan W.W. 1959, AJ 64, 432

\ref
Mould J.R., Oke J.B., \& Nemec J.M. 1987, AJ 93, 53

\ref
Mould J.R., Oke J.B., De Zeeuw P.T., \& Nemec J.M. 1990, AJ 99, 1823

\ref
Ortolani S., Renzini A., Gilmozzi R., et al.~1995, Nature 377, 701

\ref
Peebles P.J.E., \& Dicke R.H. 1968, ApJ 154, 891

\ref
Persson S.E., Aaronson M., Cohen J.G. et al.~1983, ApJ 266, 105

\ref
Puzia T.H., Kissler-Patig M., Brodie J.P., \& Huchra J.P. 1999, AJ in press

\ref
Richtler T. 1994, Reviews in Modern Astronomy, Vol.~8, ed.G.Klare, p.163

\ref
Schlegel D.J., Finkbeiner D.P., \& Davis M. 1998, ApJ 500, 525

\ref
Schweizer F. 1987, in ``Nearly normal galaxies'' ed.S.M.Faber, 
New York:Springer, p.18

\ref
Schweizer F. 1997, in ``The nature of elliptical galaxies'', ASP Conf.
Ser., Vol.~116, eds.M.Arnaboldi, G.S.Da Costa, P.Saha, p.447

\ref
Schweizer F., \& Seitzer P. 1998, AJ 116, 2206

\ref
Schroder L.L., Brodie J.P., Kissler-Patig M., et al.~2000, AJ submitted

\ref
Searle L., \& Zinn R. 1978, ApJ 225, 357

\ref
Shapley H. 1918, ApJ 48, 154

\ref
Sharples R.M., Zepf S.E., Bridges T.J., et al.~1998, AJ 115, 2337

\ref
Toomre A., 1977, in ``The Evolution of Galaxies and Stellar
Populations'', eds. B.M.Tinsley \& R.B.Larson, New Haven:Yale University
Observatory, p.401

\ref
van den Bergh S. 1982, PASP 94, 459

\ref
van Dokkum P.G., Franx M., Fabricant D., et al.~1999, ApJ 520, L95

\ref
Whitmore B.C. 1997, Whitmore, B.C. 1997, in ``The Extragalactic Distance
Scale'', Symp. Ser., Vol. 10 STScI, Cambridge University Press, p.254

\ref
Whitmore B.C., Zhang Q., Leitherer C., et al.~1999, AJ 118, 1551

\ref
Worthey G. 1994, ApJS 95, 107

\ref
Zepf S.E., \& Ashman K.E. 1993, MNRAS 264, 611

\ref 
Zepf S.E., Ashman K.E., English, J. et al.~1999, AJ 118, 752

\ref
Zinn R. 1985, ApJ 293, 424
}

\vfill
\end{document}